\documentclass[10pt,aps,prl,twocolumn,showpacs,superscriptaddress]{revtex4-1}
\usepackage{amsmath,graphicx}
\usepackage{amssymb}
\usepackage{algpseudocode}
\usepackage{bm}
\usepackage{color}
\usepackage{graphicx}
\usepackage{xcolor}

\usepackage{amsthm}
\newtheorem{mydef}{Definition}

\begin{document}
\title{Fast phase retrieval for high dimensions: A block-based approach}

\author{Boshra Rajaei$^{1,5, 6}$, Sylvain Gigan$^{3,6}$, Florent Krzakala$^{2,6}$, Laurent Daudet$^{1,4,6}$}
\noaffiliation
\affiliation{
Institut Langevin, ESPCI and CNRS UMR 7587, Paris, F-75005, 
France\\
$^2$LPS-ENS, UPMC and CNRS UMR 8550, Paris, F-75005, France.\\
$^3$Laboratoire Kastler Brossel, UPMC, ENS, Coll\`ege de France, CNRS UMR 8552, Paris, F-75005, France \\
$^4$Paris Diderot University, Sorbonne Paris Cit\'e, Paris, F-75013, France\\
$^5$Sadjad University of Technology, Mashhad, Iran\\
$^6$ PSL Research University, F-75005 Paris, France
}

  \begin{abstract} 
This paper addresses fundamental scaling issues that hinder phase retrieval (PR) in high dimensions. We show that, if the measurement matrix can be put into a generalized block-diagonal form, a large PR problem can be solved on separate blocks, at the cost of a few extra global measurements to merge the partial results. 
We illustrate this principle using two distinct PR methods, and discuss different design trade-offs. Experimental results indicate that this block-based PR framework can reduce computational cost and memory requirements by several orders of magnitude. 
 \end{abstract} 
\maketitle 

\section{Introduction}\label{sec:Introdution}  
Phase retrieval (PR) is the problem of recovering a complex-valued signal $\mathbf{x}\in \mathbb{C}^N$ from the squared magnitude $\mathbf{y} \in \mathbb{R}_+^M$ of its (possibly noisy) projections 
\begin{equation}\label{equ:pr}
\mathbf{y}=| \mathbf{H}\mathbf{x} |^2
\end{equation}
where $\mathbf{H} \in \mathbb{C}^{M\times N}$ is a known matrix called projection (or measurement) matrix. This problem arises in many digital signal processing situations, such as audio source separation, but also in physical sensing / imaging applications where designing an intensity-only detector (such as most optical sensors) is easier, faster and/or cheaper than amplitude-and-phase detectors \cite{dremeau15}. Some of these applications include X-ray crystallography \cite{harrison93}, X-ray diffraction imaging \cite{bunk07}, optical imagers \cite{walther63,liutkus14} and astronomical imaging \cite{fienup87}. 
Most PR methods have been designed for Fourier transform or i.i.d. random complex measurement matrices, but in some applications a generic solution to Eqn. \eqref{equ:pr} without specific restrictions on $\mathbf{H}$ and/or $\mathbf{x}$ may be required. Well-known PR methods include but are not limited to convex relaxation algorithms such as phaseLift \cite{candes13} and phaseCut \cite{waldspurger15}, iterative non-convex optimization algorithms such as Wirtinger flow (WF) \cite{candes15} and its truncated version (TWF) \cite{chen15}, iterative projections algorithms such as Gerchberg and Saxton \cite{gerchberg72}, Fienup \cite{fienup78reconstruction} and  variants  \cite{marchesini07,netrapalli15}, and spectral recovery method \cite{alexeev14}.  

This study investigates scalability issues for PR algorithms, as a function of  the size $N$ of the unknown signal $\mathbf{x}$. 
To reconstruct the complex signal $\mathbf{x}$ (up to a global phase) using its intensity-only projections, the size $M$ of the measurement vector should be at least $2 N$ - it has been established recently that, in a generic case, $M\geq4N$ measurements are required \cite{bodmann15} to recover a unique $\mathbf{x}$. Therefore, the amount of data that a PR algorithm has to handle, for the $\mathbf{H}$ matrix,  is at least of the order $O(N^2)$. Besides these memory requirements, the computational complexity of generic PR algorithms scales at least with the same order $O(N^2)$, and possibly worse. This can be a bottleneck for many of the above applications, such as real-time imaging. 
 
There are fundamentally two ways to alleviate these scaling issues : either by making a sparsity assumption on the unknown vector $\mathbf{x}$, or by imposing some extra constraints on the measurement matrix $\mathbf{H}$.  
In the first case, a sparsity assumption on $\mathbf{x}$ allows a reconstruction with less than $4N$ measurements, and consequently can speed-up the reconstruction. This class of algorithms are mainly referred to as compressive (or compressed) phase retrieval methods in the literature, and are mostly based on a Bayesian framework. Examples of algorithms in this category include Moravec et al. $\mathit{l}_1$-norm algorithm \cite{moravec07}, Mukherjee and Seelamantula PR method \cite{mukherjee14}, GESPAR algorithm by Shechtman et al. \cite{shechtman14} and Schniter and Rangan prGAMP algorithm \cite{schniter15}.  
In the second case, some specific classes of measurements matrices allow PR with reduced complexity, for instance Iwen et al. method based on local correlation measurements \cite{iwen15} and Zhang and Kner phase retrieval using special binary structured matrices \cite{zhang15}. However, in most of the physical scenarios presented above, the entries of the measurement matrix cannot be designed at will, as they correspond to the physical sensing process.

\medskip

In this paper, we propose a conceptually simple but remarkably effective block-based PR framework, that can be used in combination with any PR algorithm, and that not only scales up easily to high dimensions but does not impose any predefined constraint, such as sparsity, on the input signal. The constraint on the measurement matrix is that it can be put in a generalized block diagonal form, but each block may have arbitrary entries. 
This block-based phase retrieval method starts by splitting the $M\times N$ input  problem into $K$, $m_i\times n_i$ sub-problems, where $\sum_{i=0}^{K-1}n_i=N$, $m_i=\lceil \alpha n_i \rceil$ and $\alpha=M/N$. The $K$ sub-problems are then solved in parallel using any PR method. Finally, all the partial results are merged with a few extra global measurements, by applying a low-dimension global phase tuning step.  

Since, as discussed above, the memory requirements and computational complexity of PR algorithms scale at least as $O(N^2)$, breaking down the PR problem into $K$ sub-problems of size $N/K$ results in a reduction of memory/complexity requirements by at least a factor $K$ (neglecting here the cost of the low-dimensional final phase tuning step), even in single-thread mode. Furthermore, the $K$ sub-problems can here be solved in an "embarrassingly parallel" way, opening the way for further gains if multiple computing threads are available. 
It has to be emphasized that the requirement for the measurement matrix to be put into a block diagonal form is often a mild constraint  - much milder than, for instance, imposing constraints on the non-zero entries of the measurement matrix -. In fact, this approach is compatible with many of the physicals systems above ; designing the measurement matrix as block diagonal can be interpreted as the ability to probe a large object by parts, which can be controlled by the source of illumination. Our original motivation for this study arises from such a physical imaging system \cite{rajaei15}. 
More generally,  there are a number of optical imaging setups that may benefit from this approach, such as  the LED array microscope \cite{tian15}, multiple coherent diffractive imagers (CDI)  \cite{shechtman15}, and single-shot phase imaging with randomized light (SPIRaL) \cite{horisaki16}.

In summary, the main contributions of this paper are : 
\begin{itemize}
\item the presentation of a new framework for block-based PR, working with any PR algorithm, and making no assumption on the input signal. 
\item experimental results with two distinct PR algorithms, showing the computational gains for different signal sizes and choice of parameters. 
\end{itemize}

\section{Block-based PR algorithm}\label{sec: theoretical_modeling}
For ease of notation, let us consider a noiseless square root version of \eqref{equ:pr} as
\begin{equation}\label{equ:pr_noiseless}
\mathbf{y}=|\mathbf{H}\mathbf{x}|
\end{equation} 
All the equations are convertible to the general case in a straightforward way. Assume $\mathbf{H}$ follows a block structure according to the following definition which is simply an extension of block diagonal matrices to rectangular blocks. 
 
\begin{mydef}\label{def:RBD}
A $M \times N$ block matrix is called $K$-rectangular block diagonal (K-RBD) matrix iff it can be partitioned into non-overlapping $m\times n$ blocks with non-zero entries only in blocks containing $(im,in);i=0,1,...,K-1$ entries. 
\end{mydef}
Here, for the sake of simplicity, we assume equal-length blocks with $m=\frac{M}{K}$ and $n=\frac{N}{K}$ as positive integers.    Therefore, a K-RBD matrix $\mathbf{H}$ has a structure of the form 
\begin{equation}\label{equ:H}
\mathbf{H}= \left[\begin{array}{cccc}
\mathbf{H}_0 & 0 & ... & 0\\
0 & \mathbf{H}_1 & ... & 0\\
0 & 0 & \ddots & 0\\
0 & 0 & ... & \mathbf{H}_{K-1}
\end{array}\right]_{M\times N}.
\end{equation}
Note that there is no restriction on the inner structure of the $\mathbf{H}_i$ submatrices. Correspondingly, we split the input vector $\mathbf{x}$ into $K$ equal subvectors of length $n$ 
\begin{equation}\label{equ:x}
\mathbf{x} = \left[\mathbf{x_0}, \mathbf{x_1}, ..., \mathbf{x}_{K-1}  \right]^t
\end{equation}  
Using the above definitions, our block-based PR starts by solving $K$ sub-problems of 
\begin{equation}\label{equ:y_i}
\mathbf{y}_i=|\mathbf{H}_i\mathbf{x}_i|; i=0 \dots K-1,
\end{equation}
independently. This can be done by any generic PR method. Let us call the first step as the \textit{blocking step} and the resulting estimations as $\hat{\mathbf{x}}_i$. PR methods can only recover the $\mathbf{x}_i$ variables up to a global phase, which means that even under a perfect recovery assumption we have 
\begin{equation}\label{equ:xhat_i}
\hat{\mathbf{x}}_i = \mathbf{x}_i e^{j\phi_i}; i=0 \dots K-1
\end{equation}
where $\phi_i$ phase shifts are not necessarily identical. Therefore, to preserve a unique global phase all over the input signal space, the block-based PR goes through a \textit{phase tuning step}. In phase tuning, we employ an extra set of $L=\beta K$ measurements, $\tilde{\mathbf{y}}=|\mathbf{A}\mathbf{x}|$, where $\mathbf{A}$ is a $L\times N$ projection matrix, and $\beta$ is the measurement oversampling factor, typically larger or equal to 4. Note that, as opposed to the first stage,  $\mathbf{A}$ now has to get as many non-zero entries as possible, providing global information on the signal $\mathbf{x}$.  By substituting $\mathbf{x}$ from  \eqref{equ:xhat_i} and splitting measurement matrix column-wise into $K$ equal $L\times n$ submatrices, $\mathbf{A}=\left[ \mathbf{A}_0, \mathbf{A}_1, \dots , \mathbf{A}_{K-1}\right]$, we have    
\begin{equation}\label{equ:tuning}
\tilde{\mathbf{y}} = |\left[ \mathbf{A}_0 \hat{\mathbf{x}}_0, \mathbf{A}_1 \hat{\mathbf{x}}_1, \dots , \mathbf{A}_{K-1} \hat{\mathbf{x}}_{K-1}\right] \left[\begin{array}{c}
d_0\\
d_1\\
\vdots\\
d_{K-1}
\end{array}\right] |
\end{equation}
where $d_i=e^{-j\phi_i}$. This is a quick low dimension PR problem which can be solved by the same algorithm as first step, or a different one, possibly tking into account the fact that the unknown entries $d_i$ are of modulus one. Eventually, from the phase tuning output $\hat{d}=(\hat{d_0}, \hat{d_1}, \dots, \hat{d}_{K-1})$, the final estimate of $\mathbf{x}$ is 
\begin{equation}\label{equ:xhat}
\hat{\mathbf{x}} = [d_0\hat{\mathbf{x}}_0, d_1\hat{\mathbf{x}}_1, \dots, d_{K-1}\hat{\mathbf{x}}_{K-1}]^t
\end{equation} 
The proposed block-based PR algorithm is summarized in Algorithm 1. 

\begin{figure}[t]
  \centering
    \includegraphics[width=.5\textwidth]{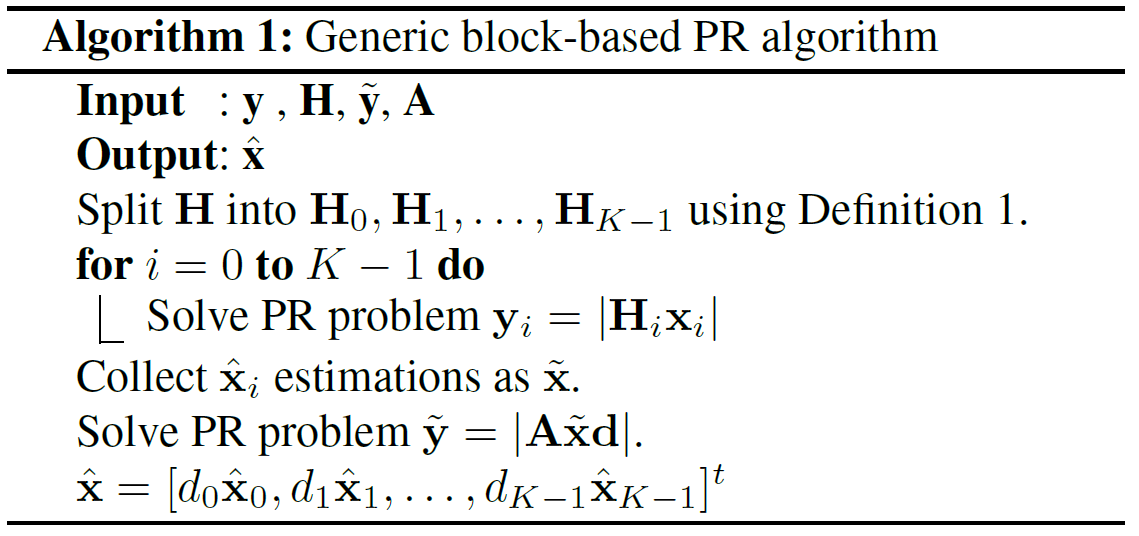}
\label{} 
\end{figure}

In the next section, we test this framework using two different PR algorithms : a non-convex minimization approach suited for gaussian entries in the measurement matrix, and a Bayesian PR algorithm designed for  binary sensing matrices. However, in general the blocking step may be accomplished by any PR method. Since the $K$ sub-problems in this stage are inherently independent (also called "embarrassingly parallel"), in a fully parallel computing configuration, the block-based PR theoretically yields at least a $K^2$ factor in computational complexity, and in single-thread sequential computing this factor is equal to $K$. One should notice that this speedup comes at the cost of a phase adjustment step. In a nutshell, assuming the computational complexity of the base PR algorithm as $O(f(N))$, the parallel block-based PR converts this order into $O(\frac{f(N)}{K^2}+f(K))$. This means that, for state of the art PR algorithms with $f(N)\sim N^2$, and by assuming significantly large $N$, $O(\frac{f(N)}{K^2}+f(K))<f(N): \forall N>K$. In addition to the computational complexity of the underlying PR algorithm, the optimal value of $K$ also depends on the number of available processing units in the blocking step. We discuss this more in the next section.
  
\section{Experimental Results} 
\label{sec: simulations} 

\subsection{Block-based PR with truncated Wirtinger flow} 

\begin{figure}[t]
  \centering
    \includegraphics[width=.5\textwidth]{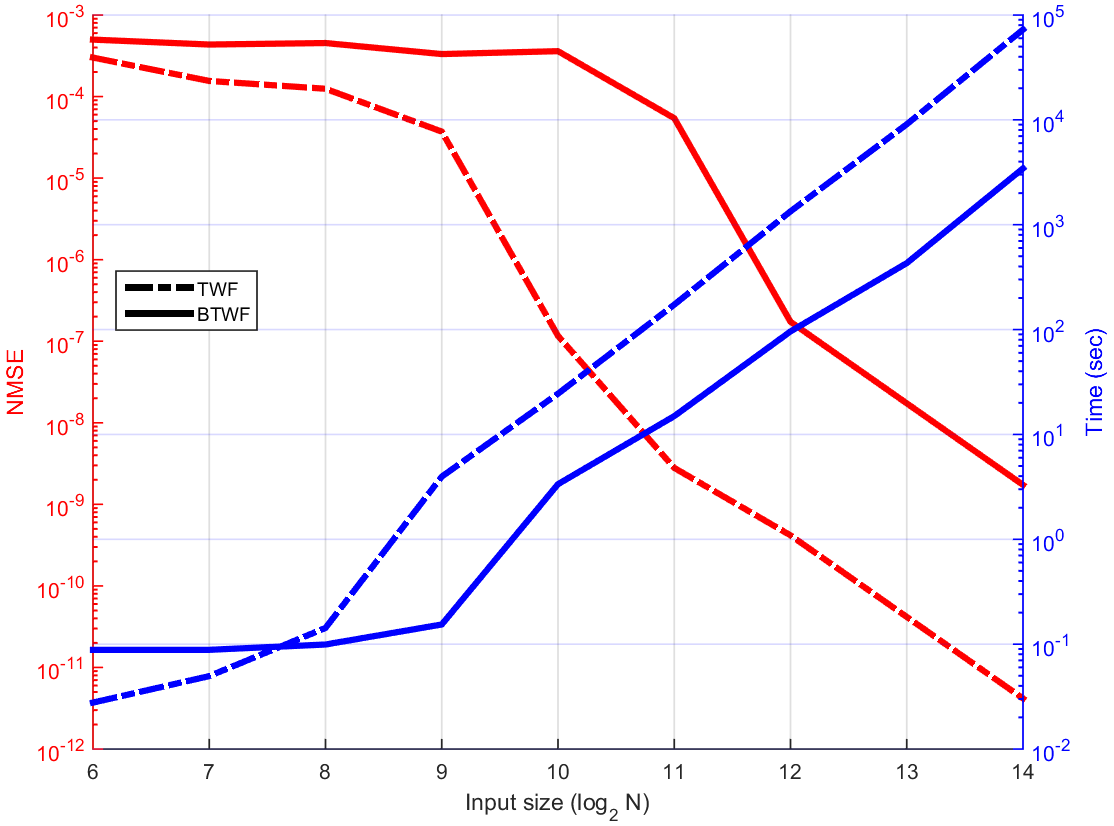}
\caption{Comparison of estimation error (in red, left scale) and execution time (in blue, right scale) between TWF (plain lines) and Block-based TWF with $K=4$ blocks (BTWF - dashed lines), as a function of the input size $N$.  }
\label{fig:BTWF} 

\end{figure}

To investigate the performance of the proposed block-based phase retrieval approach, we first employ a recent algorithm based on truncated Wirtinger flow (TWF)  \cite{chen15} to solve the PR sub-problems in the blocking step. The TWF method, currently considered amongst the state-of-the-art for generic PR,  has been reported to follow $O(MN)$ computational complexity. Since the number of required measurements, $M$, grows linearly with the number of input samples, $N$, this order actually resembles $O(N^2)$. Figure \ref{fig:BTWF} represents the effect of employing the block-based approach to improved TWF algorithm using only $K=4$ blocks. Here, the input signal, $\mathbf{x}$, and the partitions of K-RBD projection matrix, $\mathbf{H_0}, \dots, \mathbf{H_3}$, have i.i.d. zero-mean complex random Gaussian entries. Gaussian i.i.d. noise is added to the squared magnitude measurements, with SNR=30 dB. In this experiment, $\alpha=6$ and $\beta=20$. The blocking step is executed in parallel and a simple PR with alternating projections \cite{bauschke02}  is employed in the final $K=4$ dimensional phase tuning step. Each value is the average result over 100 random test inputs, using a 4 cores 3.2 GHz processor with 32 GB of RAM. The performance is measured using the normalized mean square error (NMSE) between original and estimated signals after compensating the global phase shift. The results show up to 20 times speedup with the block-based approach. It has to be noted that this comes at the price of a small loss in precision:  local measurements  carry information on a smaller number of input coefficients, and are therefore more sensitive to numerical / experimental noise. 

As mentioned in the previous section, the optimum number of blocks, $K$, depends on the computational complexity of the employed PR algorithms in the blocking and phase tuning steps, in addition to the achievable degree of parallelism.
 Clearly, by increasing $K$ and hence decreasing the block size, we have a faster algorithm in the first step -  at the cost of a more complex phase tuning step with $K$ variables.   


\begin{table}[!htbp]
\caption{Speedup factor in computation time provided by the block-based PR method, for various input variable size, $N$. For each $N$, the best number $K$ of blocks is chosen based on computation time, keeping the relative NMSE below $10^{-3}$.}
\begin{center}
\begin{tabular*}{\columnwidth}{l|lllllll}
$N$              & $2^8$ & $2^9$ & $2^{10}$  & $2^{11}$ & $2^{12}$ & $2^{13}$ & $2^{14}$ \\
\hline\hline
$K$            & 4 & 4 & 8 & 16 &  32 & 64 & 64 \\
\hline
Speed-up & 1.2 & 27 & 103 &	343 & 558 & 3398 & 9295 \\

\end{tabular*}
\label{tbl:Kstar}
\end{center}
\end{table}

Beside execution time, another important factor is the estimation error. By increasing the number of blocks in the phase tuning step,  the estimation error increases. Suppose we tolerate a $10^{-3}$ error in terms of NMSE for both the original TWF and its block-based variant. Then, Table \ref{tbl:Kstar} shows the optimal $K$ and the best speedup factor one can achieve using the block-based approach for various input size $N$. Empirically, the optimal $K$ roughy scales as $N^{0.4}$, which is close to the theoretical prediction $K^*=\mbox{argmin}_K c_1\frac{N^2}{K^2}+c_2 K^2 = c_3 N^\frac{1}{2}$, with $c_1$, $c_2$ and $c_3$ as constants.

 \subsection{Block-based Bayesian PR} 
As a second experiment, we examine the proposed block-based approach on a PR algorithm using binary $\{ 0, 1\}$ measurement matrices. In some physical situations, employing binary projections instead of random complex values makes the  measurement setup simpler. However, the corresponding ill-conditioned measurement matrices make PR more challenging, as for instance the TWF typically fails. In \cite{rajaei15}, the authors suggest a Bayesian-based PR algorithm called prSAMP - for phase retrieval swept approximate message passing. The method which originates from SwAMP \cite{manoel14} and prGAMP \cite{schniter15} algorithms, solves $\mathbf{y}=|\mathbf{H}\mathbf{x} + \mathbf{w}|^2$ problem where $\mathbf{H} \in \{0,1\}^{M\times N}$ is the known binary measurement matrix, $\mathbf{x}\in \mathbb{C}^{N}$ is the unknown complex signal and $\mathbf{w}\in \mathbb{C}^{N}$ is the (unknown) noise, assumed i.i.d. complex Gaussian. Even though the algorithm performs well for a real optical imager and strong noise conditions \cite{rajaei15}, its $O(N^3)$  computational complexity makes it impractical at high dimensions. 

\begin{figure}[t]
  \centering
    \includegraphics[width=.45\textwidth]{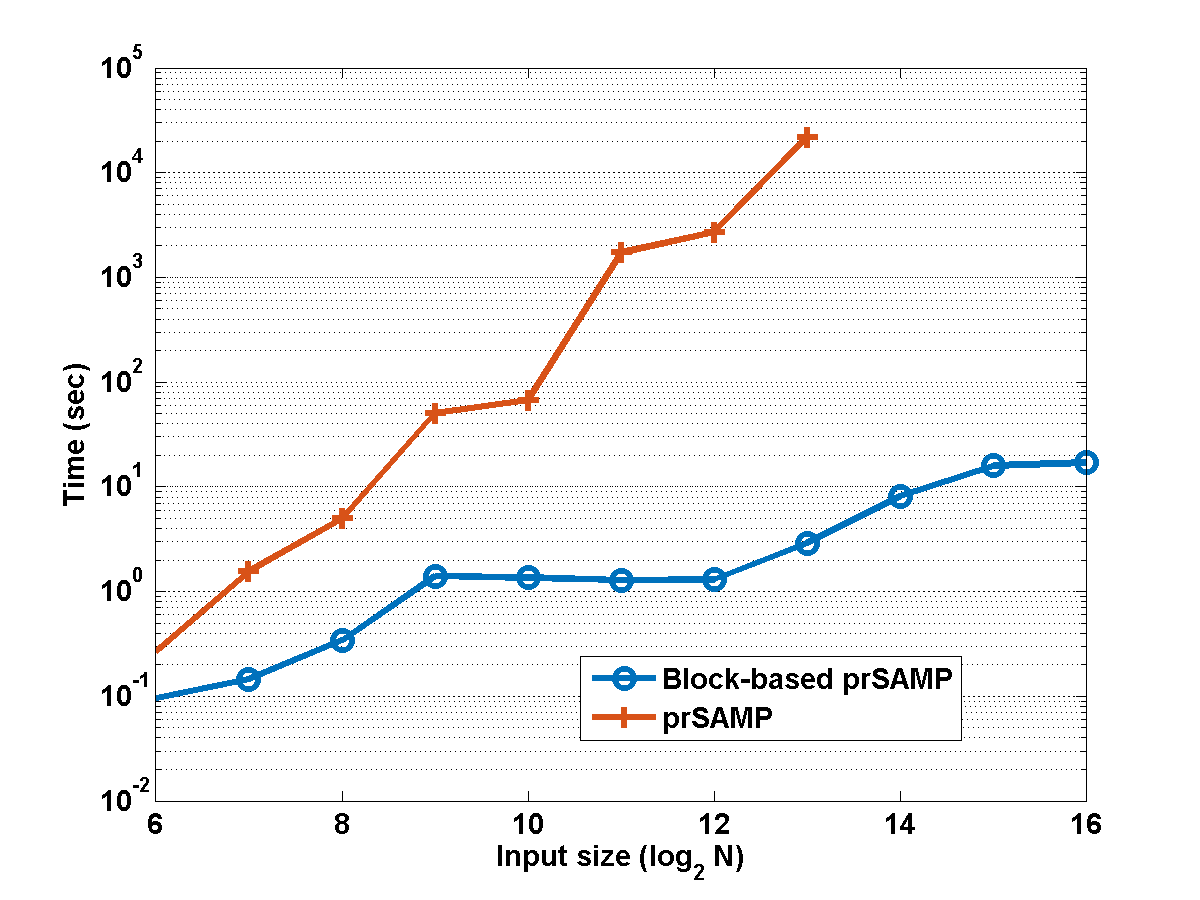}
    
\caption{Computing time (on a log scale) with the prSAMP algorithm, comparing the standard (red) and the block-based approach (blue) using the optimal number of blocks. }
\label{fig:prSAMP} 

\end{figure}

Figure \ref{fig:prSAMP} compares the execution time of the original prSAMP algorithm and its blocked version at different input sizes $N=64$ to $65536=2^{14}$,  for a comparable NMSE. The optimal number of blocks is set using the same approach as in the previous section. As expected, the block-based variant brings very significant speedups to this $O(N^3)$ algorithm. In this experiment, for instance, the block-based prSAMP approach is more than 7000 times faster than the original algorithm at $N=8192$. Beside the computational complexity, the memory requirement grows as $O(N^2)$  to store the measurement matrix, which in practice is also an important  bottleneck. For instance, at $N=2^{14}$ more than 20 GB of RAM is required to store this matrix in double precision. Due to other temporary variables for AMP messages, the original prSAMP stopped executing at $N=2^{14}$ on a computer with 32 GB RAM. In reverse, the block-based version could still be run at $N=2^{16}$ and higher.

\section{Conclusion}\label{sec: conclusion} 
We have introduced a framework for block-based PR, allowing substantial speed and memory savings for large signals. This comes of course at a price : first, it can only work if one is able to design the measurement matrix in a general block-diagonal manner - this is the case in any physical systems where one can probe the whole object by parts. Then, for a given number of measurements the approximation error is slightly increased. Finally, a small number of extra measurements is needed, but this number scales as the number of blocks, $K$, and does not depend on the signal dimension, $N$. 

Although, depending on the application, these may be seen as strong limitations, one should be reminded that, due to the fundamentally harsh $O(N^2)$ scaling laws of generic PR, using these block-based PR might not just be a matter of mere computing time : in practice, it may be the only way to achieve PR on very large signals. 

 \section{Acknowledgment}
This research has received funding from the European Research Council under the EU’s 7th Framework Programme (FP/2007- 2013/ERC Grant Agreement 
307087-SPARCS and 278025- COMEDIA) ; and from LABEX WIFI under references 
ANR-10-LABX-24 and ANR-10-IDEX-0001-02-PSL$^\star$.


\bibliography{references}
\end{document}